\documentclass[aps, prd, preprint, tightenlines, nofootinbib]{revtex4}

\begin{document}

\title{Conflict between anthropic reasoning and observation}
\author{Ken D. Olum}
\affiliation{Institute of Cosmology,
Department of Physics and Astronomy, 
Tufts University, 
Medford, Massachusetts 02155}
\email{kdo@cosmos.phy.tufts.edu}

\begin{abstract}
Anthropic reasoning often begins with the premise that we should
expect to find ourselves typical among all intelligent observers.
However, in the infinite universe predicted by inflation, there are
some civilizations which have spread across their galaxies and contain
huge numbers of individuals.  Unless the proportion of such large
civilizations is unreasonably tiny, most observers belong to them.
Thus anthropic reasoning predicts that we should find ourselves in
such a large civilization, while in fact we do not.  There must be an
important flaw in our understanding of the structure of the universe
and the range of development of civilizations, or in the
process of anthropic reasoning.

\end{abstract}

\maketitle

\section{The problem}

Where in the universe should we expect to find ourselves?  What
should be our circumstances?  Surely we should expect to find
ourselves in situations that are common for observers.
That idea is formalized by Bostrom \cite{Bostrom:book} as the
``Self-Sampling Assumption'':
\begin{quote}
One should reason as if one were a random sample from the set of all
observers in one's reference class.
\end{quote}
Bostrom's reference class includes observers who existed in the past
or will exist in the future, and I argued elsewhere \cite{Olum:2000zw}
that even observers who might have existed should be included.  But
those extensions will not be necessary here.  The problem will exist
even if we confine ourselves to those observers who exist presently.

What is the set of presently existing observers?  If we accept the
theory of inflation, then evidently the universe is infinite.
Generically, models of inflation are eternal, meaning that there are
always places in the universe where inflation is still going on
\cite{Vilenkin:1983xq,Steinhardt:1982kg,Guth:2000ka}.  Thus an
infinite amount of space is produced.  Then, since regions of the
infinite universe that are very far apart develop independently, there
will be an infinite number of regions very similar to ours and an
infinite number different in all possible ways
\cite{Ellis-Brundrit}\footnote{In fact, there will be an infinite
number identical to ours \cite{Garriga:2001ch}.}.  Thus the set of
presently existing observers is infinitely large and very
diverse\footnote{It is not well defined to choose a random item from
an equally weighted, countably infinite set.  Instead we can consider
observers in a finite but very large spherical region, and take a
limit as the region becomes infinitely large
\cite{Vanchurin:1999iv}.}.

In the infinite universe, there are civilizations like ours, but there
are also some which are much larger.  There are stars like the sun which
formed billions of years before the sun did, and thus civilizations
which are billions of years older than ours.  It is not impossible for
such civilizations to have spread widely through the universe, and so
huge civilizations exist somewhere today.

How many individuals might belong to such a civilization?  We can
consider the population to be the product of the number of star
systems colonized and the average population of each.  Estimates
(e.g., \cite{Jones:expansion}) for the time to colonize a galaxy
completely are typically much less than a billion years, so early
civilizations could easily have filled their galaxies.  Our galaxy has
about $10^{11}$ stars.  We don't know what fraction have planets
suitable for colonization, but some estimates (e.g.,
\cite{Fogg:habitable}) give on the order of 1\%.  If the $10^9$ suitable
planets each have the current population of the earth, about $10^{10}$,
then the total number of individuals will be $10^{19}$.

What fraction of the civilizations currently existing are of this
type?  It appears that our civilization is in a short interval between
coming into existence and either dying out or spreading across the
galaxy.  Most civilizations that exist elsewhere will have either died
out already or grown large, since the time that very large
civilizations endure is likely to be much longer than the time that
they take to reach that stage.  But we can be conservative and say
that the majority, say 90\%, of the currently existing civilizations
are small ones.

According to these assumptions, the galaxy-wide civilizations have $10^{9}$
times more individuals than the small ones, and are 10\% as common, so
all but one individual in a hundred million belongs to a large civilization.
Nevertheless, we do not. Anthropic reasoning predicts that we are
typical, and thus it predicts with great confidence that we belong to
a large civilization.  Using the estimates above, this prediction is
violated in a way which has a chance probability of only $10^{-8}$.

But the situation is much worse than described above.  Some
civilizations will have colonized beyond their own galaxies.  In the
extreme case, a civilization could fill a significant fraction of the
future light cone of its origin.  It might cover as much as $10^9$
cubic megaparsecs, and so fill on the order of $10^7$ large galaxies.
Presumably a small fraction of the galaxy-wide civilizations have
spread as widely as that, but if the fraction is more than $10^{-7}$,
most individuals are in those civilizations.

Similarly, some civilizations will have many more individuals at each
star, because they are using the majority of the energy from their sun
\cite{Dyson:sphere}, rather than the approximately $10^{-9}$ that is
absorbed by an earthlike planet.  Again, it might be unusual, but such
civilizations could have $10^9$ times as many individuals at each
star, and so if more than $10^{-9}$ of civilizations take this route,
they contain most of the individuals.

It is not possible to make accurate estimates of the fraction of
civilizations choosing any of these paths, but it appears that the very
largest civilizations are likely to be the dominant ones, and that
no more than one in a billion observers, and probably an even smaller
fraction than that, are in civilizations like ours.

\section{Possible solutions}

When the predictions of a theory are violated at the level of one in a
billion, the theory must be rejected. So something is very wrong with
the derivation above. Either anthropic ideas are not correct, or we
do not properly understand the structure of the universe and the
processes which enable civilizations to grow.  In this section, I consider
various possibilities for where the problem might lie.

\subsection{Anthropic reasoning is wrong}

One possibility is that the whole idea of anthropic reasoning is wrong
and should not be used to make predictions about what we observe.  But
we must have some sort of anthropic procedure or we will never make any
progress in science.  If we reject only those theories which make
our observations absolutely impossible, rather than merely unlikely,
we will never get anywhere \cite{Bostrom:location}.

\label{sec:Higgs}
For example, consider the theory that there is a Higgs particle with a
mass of 75GeV$/c^2$.  One might think, since the current lower bound
for the mass of an undetected Higgs is 115GeV$/c^2$ at the 95\%
confidence level \cite{PDBook}, that such a theory has been ruled out.  The
argument, of course, is that if the mass were as low as 75GeV$/c^2$,
many Higgs particles would have been produced in accelerators and
detected.  But since the production process is probabilistic, there
would still be some tiny chance that no such events would have
occurred, and thus, somewhere in the universe, there would still be
civilizations which have failed to detect the Higgs.  So, if you think
that we could be in any of the civilizations, rather than expecting us to
be typical, then we might be in that unlucky civilization.  Thus without
some sort of anthropic reasoning we can never rule out any theory.

\subsection{Anthropic reasoning should use civilizations instead of
individuals}

Instead of the Self-Sampling Assumption, we could use Vilenkin's
Principle of Mediocrity \cite{Vilenkin:1995ua}, which says that we are
a typical civilization among all the civilizations, rather than being
typical individuals.  In that case, large civilizations will not
receive higher weight, and the conflict above will not occur.

Could this principle be right?  It certainly is wrong if one considers
a civilization to be just an arbitrary grouping of individuals.  For
example, we're not surprised that we are in a large galaxy rather than
a dwarf galaxy, even though dwarf galaxies are more numerous, because
large galaxies have most of the stars.

Of course the division into civilizations is not arbitrary.  We act
together as a civilization when we observe the universe and discuss
the consequences of our observations.  Could that fact be relevant?
Consider the following analogy \cite{Olum:2000zw}.  One day you
receive a letter asking you to participate in an experiment.  You know
that thousands of other people will be participating, and that you
will be part of a group of either 10 or 1000 people, with an equal
number of groups of each kind.  You should think it very likely that
you will be in one of the large groups.  If you find yourself in a
small group, you should certainly give some thought to the idea that
you didn't get selected for that role by chance, or that the
experiment is not as described.  You will come to the same conclusion
even if you discuss your reasoning with the other people in your group
after the experiment starts.

\subsection{One should consider observers who live at any time}

Instead of just those observers who exist at the present moment, one
could consider all observers who have ever lived or will ever live.
That idea seems quite reasonable, but it makes things worse.  Aggressively
expanding civilizations will generally continue to expand in the
future, and so they will contain an even greater fraction of all
individuals.

\subsection{One must take account of selection bias}

Perhaps we are only concerned about the problem because of when we
live.  Maybe members of large civilizations don't think about these
problems, because they were solved by their distant forebears.  Then
only observers in small civilizations would think about them, and so
we should not reason as if we could have been in a large one.

One can get perhaps some insight into this idea by modifying the
experiment described above.  Suppose that instead of each group of
people being put together at once, some members are added and some
removed from the groups, until the final size is reached.  If you find
yourself in a small group, you will still find that very surprising,
even though the group could in principle become large later.  If you
had been added to a large group, the other members of the group might
have discussed the problem already, but that does not seem to affect
your conclusion.

Furthermore, if one excludes observers because they are not thinking
about the sizes of civilizations, then one will bring back the problem above.
Those civilizations which have already discovered the Higgs are not
presently wondering what its mass is.  But if we exclude members of
such civilizations, then we will consider ourselves typical only among
observers who have not seen the Higgs, and thus can infer nothing from
our observations.  To solve the problem would require some reason to
exclude observers in galaxy-wide civilizations from discussions of
civilization size, while not excluding those who have discovered the
Higgs from discussions of the Higgs mass.  It is hard to see what that
reason could be.

\subsection{Infinitesimally few civilizations become large}

It's clear that there are some large civilizations.  Their existence
doesn't violate the laws of physics, so they exist somewhere
\cite{Garriga:2001ch}.  But if large civilizations are infrequent
enough, most of the individuals will be in the small civilizations, as
we are.  The problem is that the frequency needs to be less than
something like $10^{-9}$, and that is hard to achieve.  Of course
there are many reasons that a civilization might remain small,
including just deciding not to grow, and there are many causes which
might bring it to an end \cite{Leslie:book}.  It might, for example, destroy
itself through nuclear war, or it might be destroyed by some natural
disaster.  It might wipe itself out unwittingly by some technological
mistake (although the arguments of the present paper should lead it to
be very careful in trying to foresee such problems). Nevertheless, it
seems hard to imagine that every civilization but one in a billion
suffers such fates.

One might think that the ``doomsday argument'' of Carter
(unpublished), Leslie \cite{Leslie:1989,Leslie:book}, Gott
\cite{Gott:doomsday} and Nielsen \cite{Nielsen:doomsday}, in
particular in the universal form discussed in \cite{Knobe:2003js},
tells us that in fact almost no civilizations are long-lived.  The
doomsday argument is controversial (e.g., see \cite{Bostrom:book} and
references therein), but even if one accepts it, there is still a
serious problem.  The doomsday argument identifies the problem as
being that civilizations are almost all short-lived, but doesn't tell
you why they are all short-lived. Something must be wrong with our
understanding of how civilizations evolve if only one in a billion can
survive to colonize its galaxy.

\subsection{The universe is not infinitely large}

Of course, the theory of inflation might be wrong.  Even if it is
correct, it is possible to have an inflationary scenario that produces
only a finite universe, although that appears to happen only in models
specially designed for that purpose.   If civilizations are extremely rare,
there might be no large civilizations in the set of civilizations that
currently exist, and so we could still be typical.

A finite universe does solve the problem, but the solution does
not seem attractive.  The evidence for inflation is good, and
inflationary models are generically eternal.  Even if not eternal,
they usually produce a universe much larger than the universe we
observe.  Thus, the density of civilizations must be remarkably small.
If 10\% of civilizations are large, there must be no more than about
10 civilizations in the entire universe to avoid the existence of a
large one.  The problem is even worse if one includes those
civilizations which have not yet arisen or grown large.

A related idea is that a true cosmological constant gives rise to a
cosmological horizon, and we should not consider places which lie
outside it.  It's not clear why we should not count observers
outside the horizon among those we could have been, but even if so the
remaining universe is quite large to contain only a few civilizations.

\subsection{Colonizing the galaxy is impossible}

It appears that many stars have planets, and that interstellar
colonization would be possible in the future with foreseeable
technology.  But it could be that some factor of which we are unaware
prevents it.  A radical version of this idea is that we are living in
a computer simulation, and the rest of the universe does not in fact
exist \cite{Bostrom:simulation}.

\subsection{We are a ``lost colony''}

It's possible that we actually do belong to a large civilization, but
are unaware of that fact.  To include that case, one should consider
the fraction of observers who {\em appear} to belong to small
civilizations, as we do.  That fraction will be infinitesimal unless a
significant share of all large civilizations are made up in
significant part of individuals unaware of the larger civilization.
It's hard to see why large civilizations should choose to divide
themselves in this way.  Even if they do so, it is very inefficient,
in the sense that the civilizations that have remained unified will
expand more rapidly than those that are divided into isolated
colonies.  Thus if any reasonable fraction of the civilizations remain
unified, they will still account for the great majority of individuals.

\subsection{The idea of ``individual'' will be different in the future}

Perhaps civilizations more advanced than ours consist of only a single
individual, or only a single individual per planet, in whatever sense
of individual is necessary for anthropic reasoning.  In that case,
even though the civilization is very widespread, the number of
individuals is small.  A similar idea is that individuals of those
advanced civilizations are so different from us that they cannot be
considered part of the same reference class, and we should not reason
as though we could have been one of them.

Both those ideas seem to have a flaw, however, in that a civilization
might decide to colonize its galaxy without any fundamental change
in the nature of its individuals. Perhaps that is unlikely, but it
doesn't seem unlikely at the level of one in a billion.

\subsection{Many factors acting together}

Perhaps the most reasonable possibility is that several of the above
factors act together.  For example, perhaps only 10\% of civilizations
survive the danger of nuclear war, of those only 10\% survive the
dangers of nanotechnology, of those only 10\% retain their
individuality, and of those only 10\% decide to colonize their
galaxies. If one can put together 9 factors at the 10\% level, then
the level of one in a billion will be reached. It seems still somewhat
of a stretch (especially as the problem is likely to be worse than one
in a billion) but perhaps the best explanation available.

\section{Conclusion}

A straightforward application of anthropic reasoning and reasonable
assumptions about the capabilities of other civilizations predict that
we should be part of a large civilization spanning our galaxy.
Although the precise confidence to put in such a prediction depends on
one's assumptions, it's clearly very high.  Nevertheless, we do not
belong to such a civilization.

Thus something must be amiss.  One possibility is that only
infinitesimally few civilizations grow to large size, although it is
unclear what stops them.  The other is that our application of
anthropic reasoning is not correct, but then what other mistakes are
we making in our use of anthropic reasoning?

\section{Acknowledgments}

I would like to thank Luca Amendola, Nick Bostrom, Milan \'Cirkovi\'c,
Jaume Garriga, Joshua Knobe and Alexander Vilenkin for helpful
conversations.  This work was supported in part by the National
Science Foundation.


\end{document}